\documentstyle[aps,epsf]{revtex}
\begin{document}

%\twocolumn 

\title{Detecting Josephson effect in the excitonic 
condensate by coherent emission of light}

\author{A. B. Kuklov}
\address{ Department of Eng. Science and Physics,
The College of  Staten Island, CUNY,
     Staten Island, NY 10314}

%\date{\today}
\maketitle

\begin{abstract}
Coherent emission 
of light by a split excitonic 
Bose-Einstein condensate -- excitonic 
Josephson junction -- can be 
extremely sensitive to the 
Josephson phase established across the junction. 
As a result of this, the emission can be 
redistributed between different directions and
even cancelled. 
The cases of the dipole- and of the  quadrupole-active 
excitons are considered.
\\

\end{abstract}
\vskip0.5 cm

Bose-Einstein condensation of low density excitons attract
much of experimental and theoretical efforts
 \cite{BEC,WOLFE,BUTOV,GOTO,SNOKE}. 
Depending on the nature of the excitons,
the excitonic condensate may display a variety
of properties specific for superfluids and superconductors.
One of these is Josephson effect. 
In Ref.\cite{AIHARA} it has been shown that an optical
analog of the Josephson effect -- optical
Josephson effect -- can be observed 
in the excitonic condensate. This effect can be 
revealed by the phase-sensitive transmition of light
\cite{AIHARA}.
 In Ref.\cite{LOZO2}
it was suggested that the Josephson current can be
detected in the case of the condensate of indirect excitons due
to spatial separation of holes and electrons. 

Reports on realization
of the excitonic condensate \cite{WOLFE,GOTO} are based
on observing a phonon assisted excitonic photoluminescence whose 
spectrum is proportional to the excitonic distribution in the energy
space. Such a method does not test directly coherence of the 
excitonic cloud. In Ref.\cite{BUTOV} the excitonic photoluminescence
from the quantum wells
subjected to magnetic field have exhibited large noise which was considered 
as an evidence of the excitonic condensation stimulated by the
field and characterized by finite size domains.  
Recently, several proposals
have been discussed 
\cite{LOZO,LAIK,MERLIN,FERNAN}. These rely on observing
light produced by the condensate of optically active excitons. 
In Ref.\cite{LOZO}, the two photon emission has been
shown to be a unique property of the condensate.
Statistical properties of the emitted 
radiation have been addressed in Refs.\cite{LAIK,MERLIN}.
The polarization of the coherent emission
from the two component excitonic condensate has been discussed
very recently \cite{FERNAN}. It has been shown that
it is sensitive to the relative phase between the components.
 
In this paper it is pointed out that 
the excitonic Josephson effect between
two confined  excitonic condensates may affect significantly the 
intensity of the light emitted by such condensates.
Thus, observing this effect can be a direct proof of the
excitonic condensation. Besides that, various 
properties of the excitonic Josephson junction 
can be tested by means of measuring the emission intensity. 

As long as excitons can transform into light,
the local polarization is proportional to the excitonic
operator $\Psi$. 
In the excitonic condensate the mean
$\psi =\langle \Psi \rangle $ 
becomes finite due to a 
formation of the off-diagonal long range order
\cite{ODLRO}. Accordingly, 
the mean electric polarization 
$\langle P \rangle \sim \langle \Psi \rangle 
= \psi \exp (-i\omega t)$ becomes 
finite, where $\omega$ is the 
frequency of the excitonic transition.
This implies that
the emission becomes coherent, as opposed to the
case when only thermal excitons emit \cite{MERLIN}. 

Here a possibility of observing a dependence of the
emission intensity on the relative phase $\varphi $
between two excitonic condensates will be analyzed.
A simple general analysis shows that this dependence should
exist. Indeed, let us assume that two identical
confined condensates are spatially separated by some 
displacement ${\bf x}_0$ and do not
overlap significantly. Then, the two-condensate wavefunction
can be represented as

\begin{equation}
\psi ({\bf x}) =\psi_1({\bf x})+
 {\rm e}^{i\varphi}\psi_1({\bf x -x}_0),
\label{1}
\end{equation}
\noindent
where $\psi_1({\bf x})$  and 
${\rm e}^{i\varphi} \psi_1({\bf x -x}_0)$
are the wave functions
of the first and the second condensates,
respectively. The electric and magnetic fields
of the coherent radiation at the excitonic frequency
$\omega $ in the far zone are given by the 
polarization
density which is proportional to 
 $ \psi $. This leads to
the differential intensity $W({\bf q}) \sim |\psi_{\bf q}|^2$
of emission of light with the wave vector ${\bf q}$,
where $ \psi_{\bf q}$ stands for the
spatial Fourier component of
$\psi ({\bf x})$. Thus, from the
representation (\ref{1}) one finds  

\begin{eqnarray}
W({\bf q})=2 W_1({\bf q})
(1+ \cos(\varphi -{\bf qx}_0)),
\label{2}
\end{eqnarray}
\noindent
where $ W_1({\bf q}) $
denotes  the intensity due to one condensate only
when the second is not present at all.
Eq.(\ref{2}) shows that the differential emission intensity of two
condensates depends on their relative phase $\varphi $. 

If $x_0$ is much smaller than the wavelength
$\lambda = 2\pi /q$ of the emitted light, one may neglect 
the term ${\bf qx}_0$ in Eq.(\ref{2}). Then, the
total intensity 
$W^{(tot)}=\int d{\bf n} W({\bf q})$ as an integral
over the $4\pi$-solid angle (${\bf n}={\bf q}/q$)
will be very sensitive to the relative phase.
Specifically,

\begin{eqnarray}
 W^{(tot)}=2W_1^{(tot)}(1 + \cos\varphi )
\label{3}
\end{eqnarray}
\noindent
where $W_1^{(tot)}=\int d{\bf n} W_1({\bf q})$.
 It can be seen that for
$\varphi =\pm\pi $ the emission can be practically 
cancelled. This is an example of the {\it dark state}
which can be realized in the excitonic Josephson junction. 

However, if the spatial separation $x_0$  of the condensates is 
comparable with or larger than $\lambda$, 
the inegration
over the total solid angle 
may 
suppress the dependence on the phase $\varphi $
in Eq.(\ref{2}). Nevertheless, as it will be seen below,  
the total intensity of emission by two condensates of 
optically active excitons can remain strongly
sensitive to the phase for large $x_0$. In contrast,
in the case of the quadrupole-active excitons
in $Cu_2O$, the dependence on the phase practically
disappears for $x_0 >> \lambda$. 
  
First, let us analyze
the case of {\it optically active} (OA) excitons
confined inside, e.g., two identical spatially separated cubes 
with the side $2a $. The excitonic
condensate is characterized by the density $n=|\psi_1|^2$
which is taken the same for both condensates. It is naturally
to consider a limit when the excitonic healing length
~$l=1/\sqrt{ na_s}$~ is much smaller than the wavelength
$\lambda$ of the emitted radiation. Indeed, for typical
achievable densities $n\approx 10^{18}$cm$^{-3}$ \cite{BEC,GOTO}
and the excitonic scattering length $a_s$ taken as 
a typical excitonic radius $\approx 10^{-7}$cm,
one finds $l\approx 10^{-6}$cm. Under this condition
the condensate density can be taken as a constant inside
the cubes. 
To specify a relative position of the cubes,
these are assumed to be identically
oriented and displaced by $x_0 > 2a$ 
along one of their 4th-fold symmetry axis,
which will be called the $x$-axis. 
The 
excitonic polarization vector ${\bf P}$
is directed along another 4th-fold symmetry axis,
which will be called the $z$-axis. Thus, the 
polarization 
density vector  $(0,0,P_z)$ can be represented as \cite{ELL}

\begin{eqnarray}
P_z= d\Phi (0)\Psi ,
\label{11}
\end{eqnarray}
\noindent
 where $d$ denotes
the interband dipole matrix element;
$\Psi $ stands for the operator of the
excitonic center of mass motion, and $\Phi ({\bf r})$
is the wave function of the relative motion of the electron and the 
hole forming the exciton. 
In the condensate phase at the temperature $T=0$
the operator $\Psi$ is
to be replaced by the condensate wave function
$\psi$ \cite{BOG}. 

Given (\ref{11}) for a pure condensate case, 
 one finds 
the differential intensity $W_1({\bf q})$ in Eq.(\ref{2}) 
as 

\begin{eqnarray}
W_1({\bf q})=W_0\left ({\sin (aq_x)
\sin (aq_y)\sin (aq_z)
\over q\sin \theta\cos \theta \sin \phi \cos \phi }\right )^2,
\label{4}
\end{eqnarray}
\noindent
where within the orthogonal coordinate system
$x, y, z$ the angles $\theta $ and $\phi $ give
the direction of the vector ${\bf q}$ so that
the components of this vector are
$q_z =q \cos \theta$, $q_x=q\sin \theta \cos \phi$
and $q_y=q\sin \theta \sin \phi$;
the introduced constant is $W_0=8c|d\Phi(0)|^2n/\pi $,
where $c$ stands for the speed of light. A simple analysis
of the angular dependence (\ref{4}) shows
that, when $qa >>1 $,
 the emission pattern is characterized by
a very anisotropic angular distribution. Almost
all the radiation occurs inside 4 narrows cones
of the angular area $\delta \approx
1/(qa)^2 <<1$. These cones are collinear with
the $x$ and $y$ axes.
Thus, employing Eq.(\ref{4}) in Eq.(\ref{2}), one can introduce
four integrated intensities 
$W_x^{(tot)},\; W^{(tot)}_{-x},\; 
W^{(tot)}_{y},\; W^{(tot)}_{-y}$ 
emitted into the corresponding 
directions $x>0,\; x<0 ,\; 
y>0, \; y<0$.  
 It is important
to note that the intensities 
$ W^{(tot)}_y=W^{(tot)}_{-y}$, emitted
perpendicular to the vector ${\bf x}_0$,
are practically insensitive to the 
phase $\varphi$. In contrast, the intensities 
$ W^{(tot)}_x,\; W^{(tot)}_{-x}$, emitted 
collinear with ${\bf x}_0$, are very
sensitive to the phase $\varphi$.
Specifically, one finds
 
\begin{eqnarray}
W^{(tot)}_x=2W^{(tot)}_y(1+\cos (\varphi - x_0q)),\quad
W^{(tot)}_{-x}=2W^{(tot)}_y(1+\cos (\varphi + x_0q)),
\label{5}\\  
W^{(tot)}_y = W^{(tot)}_{-y}=
32 \pi c |d\Phi (0)|^2 n a^2\sin^2(qa)
\phantom{XXXXXXXX}
\label{70}
\end{eqnarray} 
\noindent
in the limit $qa >>1$. Note that for $x_0q =p\pi$
where $p$ is integer,
the emission in the both directions $x>0,\; x<0$
can be completely suppressed by choosing 
$\varphi =0$ or $\varphi = \pi$ depending
on parity of $p$. For $x_0q \neq p\pi$,
the emission can be cancelled in either 
$x>0$ or $x<0$ direction. 
It is important to note that such a phase dependence
of the emission 
is a direct consequence of the excitonic condensation. 
No such an effect will occur when no condensate is
present.

The phase difference can be induced by a difference
of the 
chemical potential  created by, e.g., some external field
or due to a slight difference in the exciton number.
The condensates can also be prepared initially with 
a certain phase difference imposed by a resonant light
\cite{AIHARA}.  
If the two condensates 
are connected by a constriction,  an excitonic
flux from one condensate  to another will pass through
this narrow bridge and exhibit a sudden change of the 
velocity. The velocity of the condensate is associated
with the gradient of the phase $\varphi$. 
Therefore, the velocity change in the 
constriction will result in a jump of $\varphi$
between two condensates. As far as the total emission
intensity (\ref{5}) is very sensitive to the
phase, the phase dynamics can be studied by means of measuring
the total emission intensity. 

A special consideration should
be given to the case of 
excitons in $Cu_20$. The paraexcitons
have been reported to form
the condensate \cite{WOLFE}. In Ref.\cite{GOTO}
the condensation of orthoexcitons has been reported as
well. However,
the orthoexcitons are characterized by fast Auger
decay \cite{BAYM}. Thus, a possibility of 
achieving a coherence time long enough
to observe the Josephson effect in a
short living condensate of the orthoexcitons
is questionable.  In what follows the case
of the long living $paraexcitons$ will be considered.
 In the non-stressed
crystal no direct radiation can be
observed from the paraexcitons.
The phonon assisted emission is inherently
incoherent. However, if a mechanical stress producing
the deformation $u_{ij}$ is applied,
a quadrupole and magnetic dipole radiation can be detected
from the radiatively recombining paraexcitons
\cite{MAKAR,STRESS}.
Below it will be shown that, if the geometry of the
confinement of the two paraexcitonic condnesates is
properly chosen, 
the excitonic Josephson
effect can be revealed in this case as well.
For sake of simplicity, the cubic crystal of $Cu_2O$ will be approximated
by isotropic medium. Then, instead of Eq.(\ref{11}),
one may write for the electric polarization density 

\begin{eqnarray}
 P_i=\sum_j Q_{ij}\Phi (0)\partial_j\Psi 
\label{12}
\end{eqnarray}
\noindent
where $Q_{ij}$ is the interband quadrupole tensor
of the optical transition. 
In isotropic medium the second-rank tensor $Q_{ij}$ can
be induced 
by the deformation $u_{ij}$ so that $Q_{ij} = g u_{ij}$
in the limit of small deformations, where $g$ stands
for some constant, which can be determined from the
experimental dependence of the intensity of the 
direct recombination
emission as a function of the applied stress \cite{STRESS}.
 Let 
the deformation be along the $z$ axis,
 so that the only
non-vanishing component is $u_{zz}$.
Thus, again, as in the case of the OA excitons,
the polarization is along $z$-axis
($P_z=gu_{zz}\Phi (0)\partial_z\psi$). Now, however,
this polarization is concentrated very close
to the boundaries where the gradients of the
excitonic condensate are the largest.
Because of this, the emitted radiation becomes
more isotropic than in the case of the OA excitons
considered above. Consequently, the term ${\bf qx}_0$
under the $\cos $ in Eq.(\ref{2}) will 
practically erase the
phase dependence unless $x_0$ is less than $q=2\pi / \lambda$
(so that Eq.(\ref{3}) can be employed).
Thus, the considered above geometry of the two cubes 
is not appropriate for the paraexcitons.
Instead, it is reasonable to consider a different arrangement.
Specifically, let us consider a slab whose width 
along the $z$-axis is $4z_0 << \lambda$, and whose base is 
a square with the side $2a >> \lambda$ oriented along
the $x$, $y$ axes. This slab is divided
in half by the plane perpendicular to the $z$-axis. The Josephson 
phase $\varphi$ is assumed to exist between these two halves. 
 Given the representation (\ref{12}), calculations of the emission 
intensity in this case show that 
the factor $W_1$ in Eq.(\ref{2}) 
is to be replaced by

\begin{eqnarray}
W^{(p)}_1({\bf q})=W^{(p)}_0\left( {\sin (aq_x)
\sin (aq_y)\sin (z_0q_z)
\over \sin\theta\sin\phi\cos\phi}\right)^2
\label{7}
\end{eqnarray}
\noindent
where $ W^{(p)}_0=(16/\pi) c|Q_{zz}\Phi (0)|^2n $
and again 
$q_z =q \cos \theta$, $q_x=q\sin \theta \cos \phi$
and $q_y=q\sin \theta \sin \phi$.
It can be seen that even in the limit $qa >> 1$ the radiation
is no more concentrated in the plane $ \theta = \pi /2 $.
Instead, it is spread with respect to the axial
angle $\theta $ around the directions 
$\theta = \pi /4,\; 3\pi /4 $. 
However, the radiation is  still 
very anisotropic with respect to the polar angle $\phi$.
It is equally emitted along the directions $\phi = 0,\; \pi,\; \pm \pi/2$
into a very small
angle  $\delta \phi \approx 1/qa <<1$.
Let us calculate the integrated intensity $W^{(p)}$
emitted into one of these directions (e.g. $\phi = 0$)
by means of
integrating Eq.(\ref{2})
over $\int^1_{-1} \; d\cos \theta 
\; \int^{\delta \phi}_{-\delta \phi}\; d\phi $~,
with the replacement
$W^{(p)}_1({\bf q}) \to W_1({\bf q})$ performed. 
This yields

\begin{eqnarray}
W^{(p)}=16\pi cnq^3z_0^2 a
|Q_{zz}\Phi (0)|^2
(1 + \cos \varphi )
\label{8}
\end{eqnarray}
\noindent
in the limit $z_0 << \lambda$. 

Now let us discuss a role of the {\it thermal component}.  
The intensities (\ref{5}), (\ref{70}), (\ref{8}) of the coherent
emission are not proportional to the total volume of the condensates.
In the case of the OA excitons the intensity (\ref{5}),  (\ref{70})
scales as
$\sim a^2$ and for the paraexcitons the intensity (\ref{8})
is $\sim a$ in the limit $a\to \infty$. In contrast to this,
the intensity of the incoherent emission $W^{(in)}$
due to the 
thermal excitons should be proportional to 
the total volume occupied by the excitons. Specifically,
 $W^{(in)}\sim a^3$ for the OA excitons,
and $W^{(in)}\sim z_0a^2$ in the case of the paraexcitons. Thus,
one may expect a masking of the coherent emission by the
thermal component. 

However, it should be noted that, while the incoherent
radiation is almost isotropic, the coherent emission 
is very concentrated in certain directions. Therefore, 
a comparison should be performed between the intensities
integrated over the small solid angles,
to where the most of the coherent radiation is 
emitted. 
In the case of the OA condensate,
the integrated incoherent emission intensity is $\sim a^3
/(qa)^2 \sim a$, where it was taken into account that
the coherent emission is concentrated into the 
cones of the angular area $\approx 1/(qa)^2$. Accordingly, 
the ratio of the incoherent to the coherent integrated
intensities is $\sim 1/a \to 0$ in the limit $a\to \infty$.
Thus, the coherent radiation should always be resolved
for large enough values of $a$.

A different situation occurs for the paraexcitons.
The angular area of the coherent emission "cones" is $
\delta \phi \approx \;1/qa$. Thus,
the integrated
incoherent intensity is $\sim a^2/qa \sim a$,
 and the ratio of it 
to the intensity (\ref{8}) is 
idependent of the largest size $a$ of the system.
In what follows the intensity of the 
incoherent emission will be calculated for the case
of the paraexcitons, and the condition insuring
that the coherent effect can be resolved will be derived.
 
In order to find the incoherent intensity,
one should employ the representation (\ref{12})
for the polarization.
At $T\neq 0$, the operator
$\hat{\Psi}=\psi + \psi '$ contains the condensate part
$\psi $ and the normal part $\psi '$. The differential
intensity of the incoherent emission at the
wave vector ${\bf q}$ is $W^{(in)}({\bf q})\sim
\langle \psi '^{\dagger}_{\bf q}\psi_{\bf q}'\rangle$
where the averaging is performed
over the thermal ensemble.
The average can be calculated by employing
the Bogolubov method \cite{BOG}. Specifically,
the thermal part is represented as

\begin{eqnarray}
\psi '=\sum_k (u_ka_k +v_k^*a^{\dagger}_k),
\label{13}
\end{eqnarray}
\noindent
 where
$u_k$ and $v_k$ stand for the Bogolubov
amplitudes: $a_k$ destroys
 and $a^{\dagger}_k$ creates a normal excitation
at the state $k$ characterized by the energy $E_k$. 
Here the quasi-particles can be characterized by
the quasi-momentum ${\bf q}$ inside the slab.
 Thus, the index
$k$ must be replaced by ${\bf q}$ in Eq.(\ref{13}). 
A direct conversion of the excitation with
given ${\bf q}$ into a photon with almost the same
${\bf q}$ (in the limit $qa>>1$) is considered. In the 
contribution due to the normal component, the
only non-zero means are $\langle a^{\dagger}_{\bf q }
a_{\bf q}\rangle$ which are the population factors
of the quasi-particles. 
Thus, calculating the emission intensity into a
unit solid angle due to the
thermal paraexciton, one finds 

\begin{eqnarray}
W^{(in)}({\bf q})={c \over 32\pi}
|Q_{zz}\Phi (0)|^2 V\sin^2(2\theta) {q^6 (|u_q|^2+|v_q|^2)
\over \exp (E_q/T)-1},
\label{14}
\end{eqnarray}
\noindent
where $V=16z_0a^2$ stands for the total
volume of the excitonic cloud;
 the quantum depletion term $\sim |v_q|^2$, which
is finite at $T=0$, has been neglected. In Eq.(\ref{14})
the amplitudes and the energy are \cite{BOG}

\begin{eqnarray}
|u_q|^2={\varepsilon_q+g_{ex}n+E_q \over 2E_q},\quad
|v_q|^2={\varepsilon_q+g_{ex}n-E_q \over 2E_q},
%\label{15} \\
\quad E_q=\sqrt{\varepsilon^2_q+2g_{ex}n\varepsilon_q}.
\label{16}
\end{eqnarray}
\noindent
In here $\varepsilon_q=q^2/2m$ denotes 
 kinetic energy of a free exciton, and $g_{ex}=4\pi a_s/m$ stands
for the two-body excitonic interaction constant
expressed in terms of the scattering length $a_s$
(in units $\hbar =1$ ). 
In Eq.(\ref{14}) the main term proportional to the
total volume of the system is selected.
In the limit $n\lambda^3>>1$ and for
the gas parameter $na_s <<1$, 
one
can safely replace the Bose factor 
in Eq.(\ref{14}) by $T/E_q$ unless $T$ is
extremely low. Here it is assumed that 
$T$ is of the order of the 
temperature $T_c\approx 3.3n^{2/3}/m$ 
of the excitonic Bose-Einstein condensation
(for one component
excitons) for typical densities $\sim 10^{18}$cm$^{-3}$.
Under these circumstances
 the expression (\ref{14}) is weakly sensitive to
the interaction constant $g_{ex}$ in the limit $E_q << T$.  
Specifically, it can be seen that $W^{(in)}$
changes by only a factor of 2, while $g_{ex}$ varies from zero
to such values that the excitonic spectrum becomes
sound like $E_q=\sqrt{g_{ex}n/m}\; q$. In the latter 
situation the $q$-dependent factor in Eq.(\ref{14})
becomes $q^6T/2\varepsilon_q$, that is independent
of $g_{ex}$. In the opposite limit ($g_{ex}\to 0$),
this factor is $q^6T/\varepsilon_q$. 

The total intensity
of the incoherent radiation emitted into the narrow 
solid angle $\delta \phi $,
 to where the most part of the coherent radiation is
collected, is ~$\approx W^{(in)}({\bf q})/qa$. Thus, 
the ratio of the integrated incoherent intensity to
that given by Eq.(\ref{8}) (for $\varphi =0$) is 
 
\begin{eqnarray}
\xi \approx  0.02{ T\over  T_cn^{1/3}z_0}.
\label{9}
\end{eqnarray}
\noindent
The coherent effect can be well resolved if $\xi <1$. 
It can be seen that for 
 not very small $z_0$, one obtains $\xi <<1$ as long as
$T<T_c$. Indeed,  
making an estimate of $\xi$ for  
the typical values of the wavelength
$\lambda \approx 600$ nm, the density $n=10^{18}$cm$^{-3}$,
the temperature $T=T_c/2$ and $2\pi z_0/\lambda =1/3$,
one finds $ \xi \approx 10^{-2}$. 
Let us also estimate the absolute intensity 
of the coherent emission from the paraexcitonic condensate.
To this end, the following maximum value of 
$Q_{zz}$ is chosen
$Q_{zz}\approx ea_B^2$, where $e$ stands for the electron 
charge and $a_B=0.5 \AA$ denotes Bohr's radius. 
$|\Phi (0)|^2 = 1/\pi a_{ex}^3$ \cite{ELL} is expressed in terms
of the excitonic radius $a_{ex}\approx 10\AA$. Thus,
for the chosen above values, one arrives at the
intensity (\ref{8}) as $W^{(p)}\sim  10^{-5}$W
for $a/\lambda =10^3$. It is worth noting that the intensity of
the coherent emission from the
condensates of the OA excitons is a factor of $\sim d^2a/q^3z^2_0Q^2_{zz}$
larger than the above estimate. Depending on the choice of 
$d$ which can be varied by external fields \cite{FUKUZAWA},
this factor can enhance the above estimate by several orders of magnitude.  

In summary, the coherent 
radiation from two excitonic condensates
can be 
sensitive to their relative phase. In the
case of the OA excitons,
the emission
can be totally suppressed or significantly 
redistributed in the opposite directions by 
means of imposing appropriate values of the phase.  
 In the case of  
the paraexcitons in $Cu_2O$ under mechanical
stress, 
the dependence of the emission on the phase
can be also very distinct, if the two condensates
form a double layer whose thickness is less than
the wavelength of the emitted radiation, 
 and if the mechanical stress
is applied perpendicular to the plane of the layer.  
 
\acknowledgments

I am grateful to J. Ketterson and 
Yi Sun for pointing out to Ref.\cite{GOTO}. I am also 
grateful to J. Fernandez-Rossier for turning my attention
to Ref.\cite{AIHARA} as well as for sending me the preprint
\cite{FERNAN}.

\end{document}